\documentclass{ifacconf}

\usepackage{graphicx}      
\usepackage{natbib}        
\begin{document}
\begin{frontmatter}

\title{Fundamental elements in the development of effective crowd control strategies} 

\thanks[footnoteinfo]{This work was financially supported by the JST-Mirai Program Grant Number JPMJMI20D1 and the JSPS KAKENHI Grant Numbers JP20K14992, JP21H01570 and JP21H01352.}

\author[Komaba]{Claudio Feliciani} 
\author[Komaba,Hongo]{Daichi Yanagisawa} 
\author[Komaba,Hongo]{Katsuhiro Nishinari}

\address[Komaba]{Research Center for Advanced Science and Technology, The University of Tokyo, 4-6-1 Komaba, Meguro-ku, Tokyo 153-8904, Japan}
\address[Hongo]{Department of Aeronautics and Astronautics, Graduate School of Engineering, The University of Tokyo, 7-3-1 Hongo, Bunkyo-ku, Tokyo 113-8656, Japan}

\begin{abstract}                
In this work, we present typical challenges encountered when developing methods for controlling crowds of people (or animal swarms). We discuss which elements shall be considered and the role they play to achieve a robust control in a variety of conditions. In particular, four different studies are reviewed, each of them investigating in detail important elements encountered in crowd steering and control. More specifically synchronization, compliance, crowd (or swarm) density and human perception are studied showing the role they play in combination. Ultimately, the success of a control strategy is determined by carefully considering the effect each element has on individuals, but also on the interactions between them, leading to the creation of a collective behavior. We will also highlight the importance of psychological and cognitive factors when dealing with human crowds, hinting at the fact that automatic control systems may achieve optimal performance, but may be not necessarily well perceived by people in terms of comfort. The discussion aims at showing recent trends and potentialities of crowd control systems, but should also warn on the risk in choosing a solution prioritizing optimization toward people's safety or comfort.
\end{abstract}


\end{frontmatter}

\section{Introduction}
With the rapid improvements of sensing technology it is now possible to collect a large number of information on traffic and process it in real-time. Cameras are often used to extract information from videos to detect and track vehicles, but pedestrian detection is also possible and applied for multiple purposes. Because cars are easier to detect and they move along precisely defined lanes, vehicle traffic control is also easier to implement. Dynamic Speed Limit (DSL) (\cite{Lu2014}) is a common method to adapt speed limit on traffic conditions, thus allowing to maximize the flow along a specific route without the need to physically introduce modifications. \par
The availability of data on pedestrian traffic and crowd motion is now also paving the way to create automatic control systems targeting humans moving on foot. Crowds have been already controlled for decades or even longer and the concept of crowd control is not new. For example, the number of open registers is increased at supermarkets when the number of people waiting rises and is decreased when deemed sufficient considering the inflow of customers. However, at the present, crowd control is normally performed by humans. In infrastructures accommodating large crowds (stadium, train station, airport, etc.) the control task is typically performed in the control rooms, where hundreds of cameras are visioned and security staff is instructed on where to move and how to guide people inside the venue (\cite{Feliciani2021}). Using machine learning and computer vision, it is now possible to digitize the information available to people working in the control room thus potentially allowing an automatic control of crowds. LiDAR or BLE/WiFi scanners can also collect complementary data potentially useful for the purpose of crowd control (\cite{Teixeira2010}). \par
But controlling people is not like controlling machines and their motion is not bounded to lanes like in the case of vehicles. For these reasons, the automatic control of crowds is a challenging task. Nonetheless, the number of studies on this subject is increasing (\cite{Molyneaux2021,Lopez2021,Lopez2022}) and automatic control systems targeting crowds may appear in the near future. \par
In this work, we discuss some of the challenges found when steering crowds (or swarms in general) with a particular emphasis on methods to change their behavior collectively. We identify fundamental elements which could determine the success of a control system and should therefore be considered in detail when developing one.

\section{Control mechanisms and effectiveness on collective systems}
In the following, we present the results from four different studies, each focusing on different (but related) challenges encountered when trying to change the behavior in humans and animals showing a collective organization.

\subsection{Synchronization and crowd density}
Pedestrians are bipedal creatures, hence, their walking velocity is determined by two factors, i.e., step size and step frequency. In high density conditions (i.e. a packed crowd), both step size and step frequency decrease, so that the pedestrian flow decreases. The decrease of step size is inevitable because there is no space for a large step in the high density condition. However, step frequency can forcibly be retained by control. \par
We conducted experiments on pedestrians in a single-file circuit where overtaking is prohibited. In the first condition, pedestrians walked normally (baseline). In the other condition, pedestrians were asked to walk with the sound of constant rhythm (70 beats per minute (BPM)) from an electric metronome (rhythm condition). The step frequency in normal walking is about 100 BPM, therefore, 70 BPM is much slower than normal walking. Accordingly, at low densities, the flow (or speed) of people in the 70 BPM rhythm condition was smaller than the baseline condition. However, in the high density condition, the result reversed, i.e., the flow in the rhythm condition became larger than the baseline condition (see table~\ref{tab:rythm_exp}).

\begin{table}[hb]
\begin{center}
\caption{Effect of rhythmic sound on pedestrians walking in a single-file circuit. The sound is an effective steering mechanism to improve flow at high density when the right rhythm is chosen. Refer to \cite{Yanagisawa2012} for details.}\label{tab:rythm_exp}
\begin{tabular}{ccc}
Condition 					& Density								& Flow 					\\
\hline
Baseline (no sound) & Low (0.47 m$^{-1}$)		& 0.41 s$^{-1}$	\\
\textbf{Baseline (no sound)} & \textbf{High (1.86 m$^{-1}$)}	& \textbf{0.15 s$^{-1}$}	\\
Rhythm (70 BPM) 		& Low (0.47 m$^{-1}$)		& 0.28 s$^{-1}$	\\
\textbf{Rhythm (70 BPM)}			& \textbf{High (1.86 m$^{-1}$)}	& \textbf{0.32 s$^{-1}$}	\\
\hline
\end{tabular}
\end{center}
\end{table}

We consider that this reversal is due to the two factors. The first one is the retention of the walking frequency. By walking with the sound of a constant rhythm, the walking frequency of the pedestrians was retained at 70 BPM, which is much smaller than people walking in unimpeded conditions, but might be larger than walking in a packed crowd. The second factor is synchronization. In the high density situation, some pedestrians moved promptly with a small space, while the others did not move and waited for enough space to make one large step. The rhythm removed this heterogeneity in pedestrians’ movement and synchronized them. If the movement of pedestrians is synchronized, one can anticipate the movement of the predecessors and step forward without hesitation. We speculate that such movement by synchronization also contributed to the increase in flow.

\subsection{Compliance and information sharing}
Compliance is another important element to consider when developing strategies aimed at crowd control. In the study considered above, participants were asked to follow specific instructions and therefore behaved accordingly. However, this is a rare occurrence in pedestrian crowds and complete compliance is almost never reached. It is therefore necessary to understand how the degree of compliance can affect complex systems and how it will affect the dynamics at the collective level. Although it is possible to achieve different levels of compliance experimentally by asking only some participants to follow the instructions, there are practical limitations on this approach. Simulation is therefore an effective tool to study the effect of compliance from a theoretical perspective and yet get an insight on the role it plays.

\begin{figure}
\begin{center}
\includegraphics[width=3.5cm]{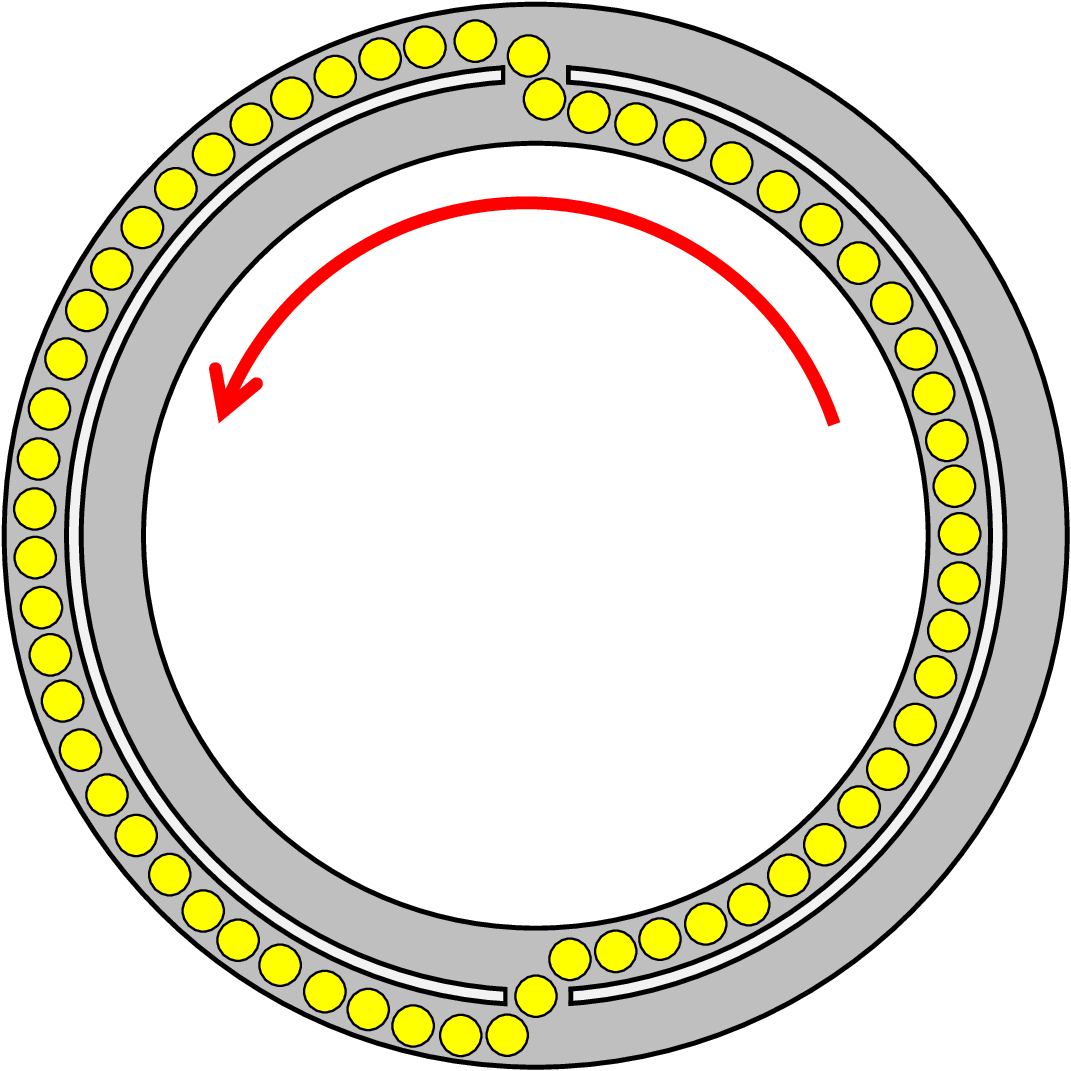}
\includegraphics[width=3.5cm]{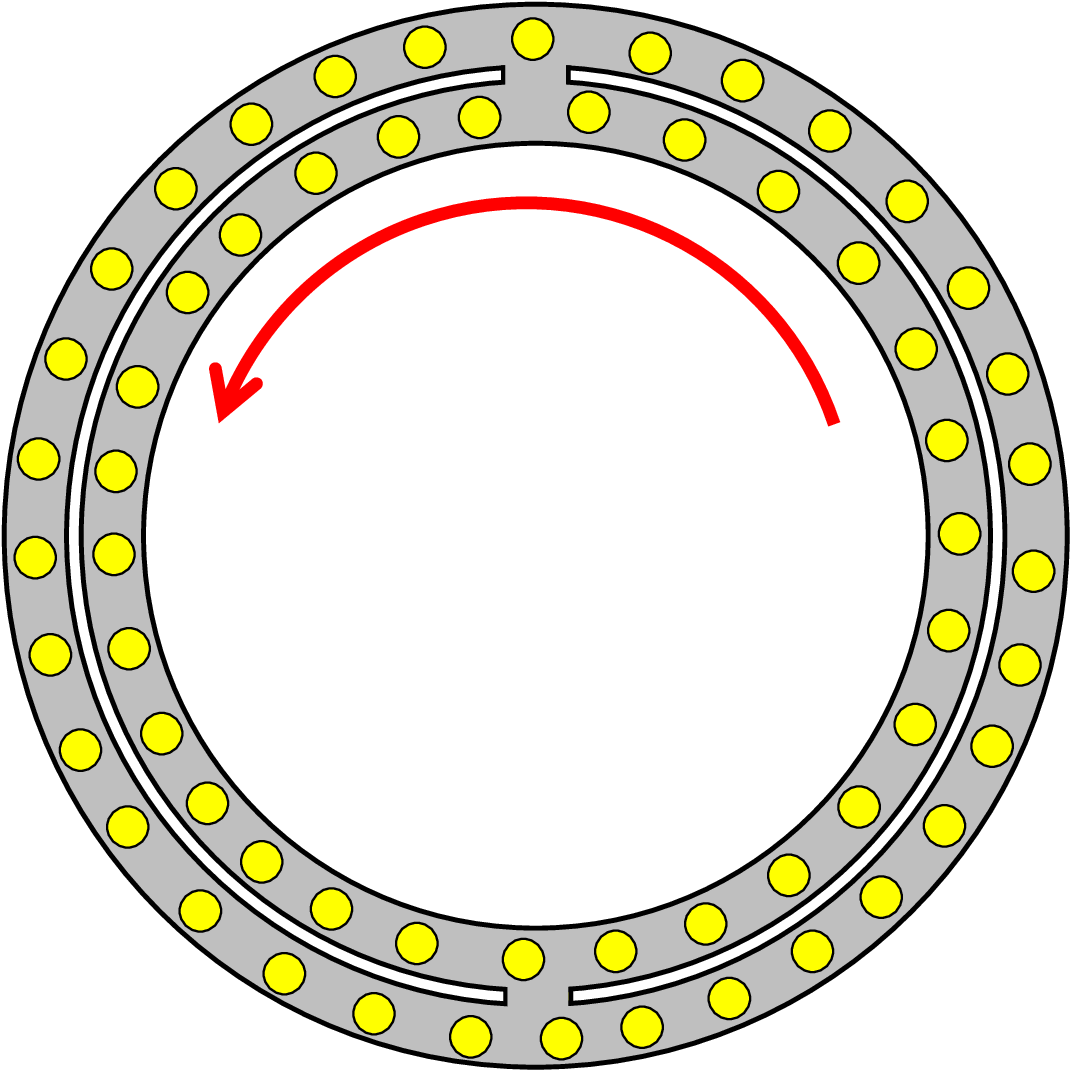}    
\caption{The two-lane setup used to study compliance in a numerical simulation; the arrow indicates the direction of motion. The left side shows an example for a sub-optimal control strategy (worst-case). If only the state of the incoming traffic is used and information not shared, the control system of each switch will guide people to the other lane in an attempt to reduce congestion. Optimal control is only reached when the global state is known and both switches take decisions by coordinating between themselves.} 
\label{fig:course_sim}
\end{center}
\end{figure}

Here, we will study the effect of compliance in a loop configuration composed of two lanes moving in the same direction (see Fig.~\ref{fig:course_sim}). Along the course it is possible to change lanes only in two locations, either from the inner to the outer lane or in the opposite direction. The number of agents (pedestrians) to be considered in the system is set to create a condition in which the maximum flow is reached when both lanes have the same number of people. When all the people move along a single lane (or in a ``snake'' configuration like on the left of Fig.~\ref{fig:course_sim}), the high density restricts motion and sub-optimal conditions are obtained. \par
We will consider three control strategies used to direct people in the switch location. The worst-case scenario makes use of local information (traffic condition before the switch) and no communication between both switch locations happens. In the best-case scenario global information is used (the whole course) and coordination occurs between both switches. Finally, a realistic scenario is considered where decisions are based on local conditions, but both switch locations share their information. Further, we assume that agents have no idea on which lane is faster and need therefore to (dis)trust indications provided at the switch locations. In short, they can only opt for following the orders given them or neglect them, with the ratio between those complying and the total defined as the compliance ratio.

\begin{figure}
\begin{center}
\includegraphics[height=4.5cm]{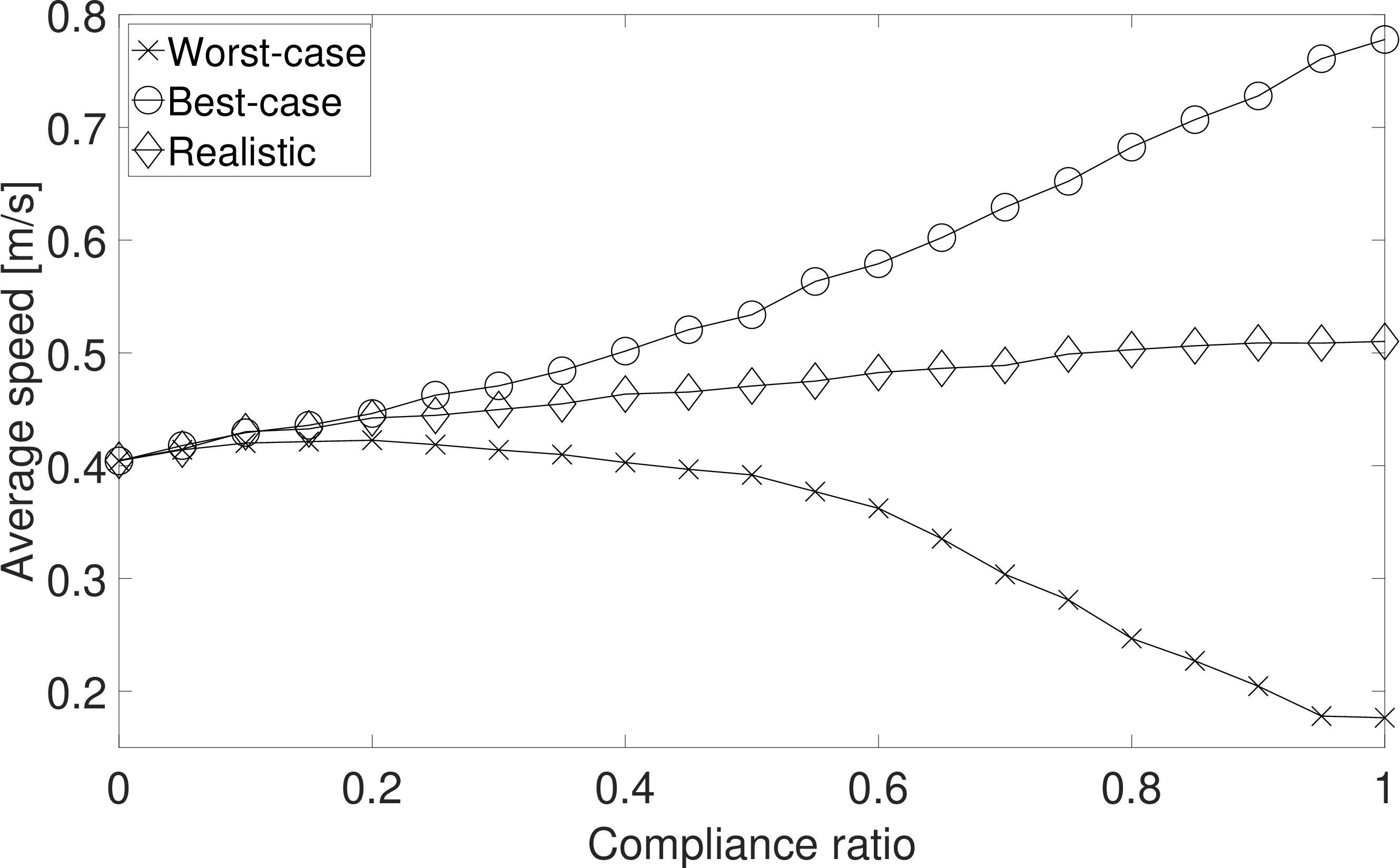}
\caption{How control strategy and compliance ratio affect the system performance (people speed here) in a numerical simulation (refer to \cite{Feliciani2018} for details).} 
\label{fig:control_sim_results}
\end{center}
\end{figure}

Results of the simulation are shown in Fig.~\ref{fig:control_sim_results}. The outcome clearly shows that both strategy chosen and compliance have a strong effect on pedestrian flow, with the control method playing the most important role in this situation. In fact, when the strategy employed is a complete failure, the negligence of people may be actually beneficial for the overall system showing that, in some conditions, low compliance may actually help improve performance.

\subsection{Perception and the human factor}
\begin{figure}
\begin{center}
\includegraphics[height=3.5cm]{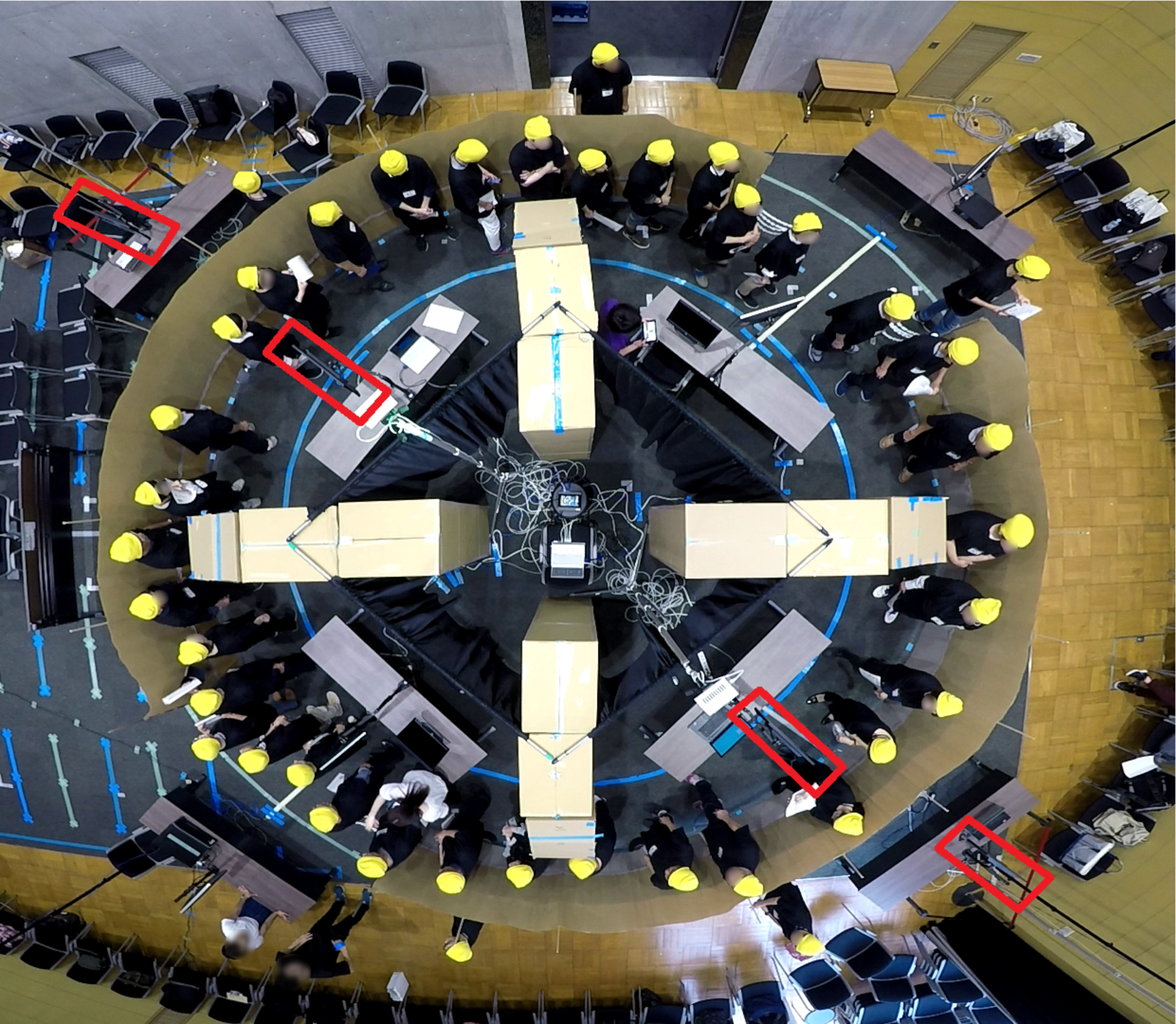}
\includegraphics[height=3.5cm]{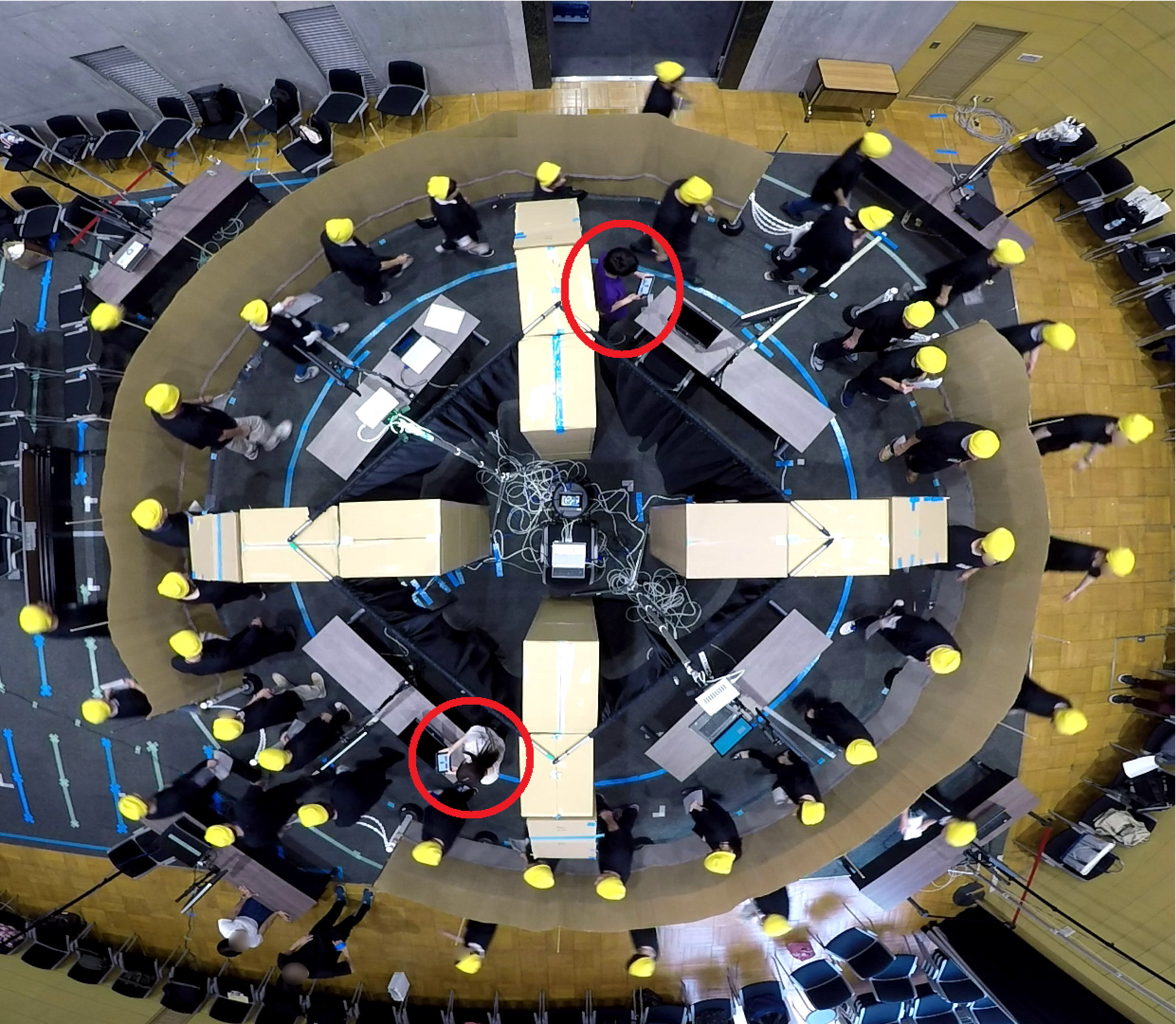}
\caption{The course used to test experimentally the influence of information provision while performing crowd control. The monitors used to show lane speed and inform pedestrians are indicated in red on the left. Guidance staff, highlighted on the right, was equipped with a remote controller displaying arrows in the switch location and guiding people to the inner or outer course. Images were partially modified in this representation to allow an easier recognition of participants' caps. Refer to \cite{Feliciani2020} for details.} 
\label{fig:control_exp}
\end{center}
\end{figure}

Although simulation can help studying some aspects related to crowd control more systematically, human behavior is best studied experimentally. Therefore, a setup similar to the one considered in the previous section was also investigated experimentally. A schematically identical course was prepared with the aim to recreate the conditions numerically studied earlier (see Fig.~\ref{fig:control_exp}). 

The main scope of the experiment was to investigate which kind of information or guidance strategy can help people move faster overall and also study the psychological perception of different control methods. Consequently, in two conditions, real-time speed information (measured using special sensors) for each lane was provided using monitors. People walking along the course could use the speed of each lane as a reference for their decision (to change lane or keep moving on the same one). In another two conditions staff on-site guided people indicating whether they should move to the inner or outer course in each switch location. In the experiments, compliance was simply varied on a two level scale, letting people decide by themselves whether they should change lanes or not or asking them to strictly refer to information or guidance provided to them (i.e. choose the fastest lane or follow guidance orders). Finally, participants were asked to rate the comfort level of each experiment.

\begin{table}[hb]
\begin{center}
\caption{Results for the double-loop experiment where participants have to move over a circuit formed by two lanes connected in two points. Comfort was obtained by asking participants to rate each procedure on a scale from 1 to 7. Refer to \cite{Feliciani2020} for details.}\label{tab:info_exp}
\begin{tabular}{ccccc}
Real-time 	& Human			& Compliance	& Flow 				& Comfort \\
speed	data	& guidance	&							& [s$^{-1}$]	& [1--7]	\\
\hline
\textbf{Yes} 	& \textbf{No} 		& \textbf{Free choice}	& \textbf{0.85}		& 3.88	\\
Yes 	& No 		& Must obey		& 0.77		& 3.38	\\
\textbf{No} 		& \textbf{Yes} 	& \textbf{Free choice}	& 0.79		& \textbf{4.44}	\\
No 		& Yes		& Must obey		& 0.77		& 4.18	\\
\hline
\end{tabular}
\end{center}
\end{table}

Results presented in table~\ref{tab:info_exp} show that there is no control strategy that is both efficient and well-perceived. The best performance in terms of flow was reached when people were informed, but they prefer human guidance instead of sole information when asked about the perceived comfort. It is also interesting to note that people preferred being controlled by human staff even when they had to follow their decisions (the comfort score is always higher in case of human guidance regardless of compliance), indicating that when psychological effects are taken into account crowd control is not an easy task.

\subsection{System density and local interactions}
Finally, we will consider a swarm of crabs to study the effect of density in conditions difficult to study with human participants. A specific species of crabs (the so-called soldier crab), known to form large swarms in nature, is used to investigate whether steering efficiency is affected by density in non-unidimensional conditions. In fact, in all scenarios considered above, only unidirectional geometries were studied, which are sufficient for a first hand analysis, but quite different from ecological environments.

\begin{figure}
\begin{center}
\includegraphics[height=3.5cm]{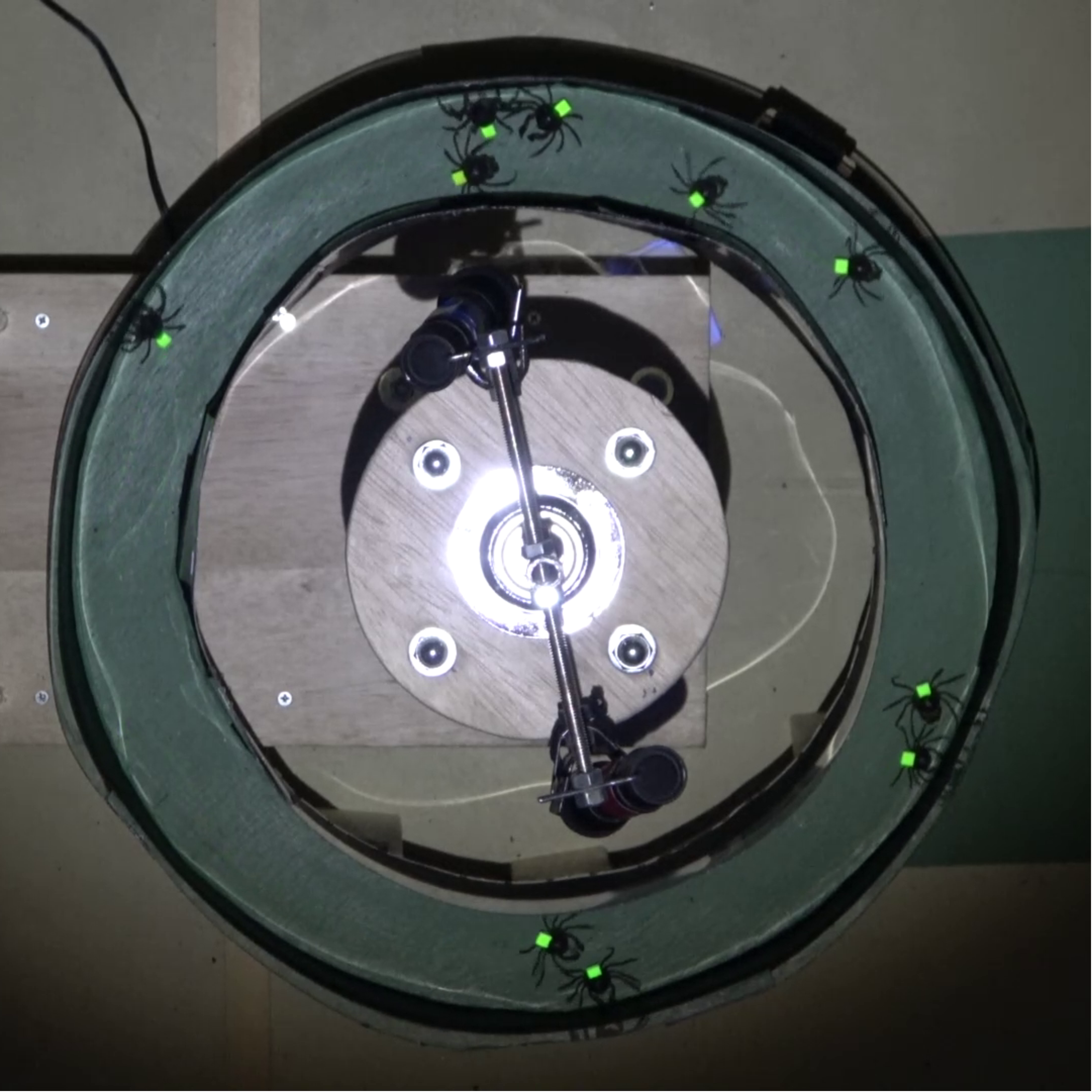}
\includegraphics[height=3.5cm]{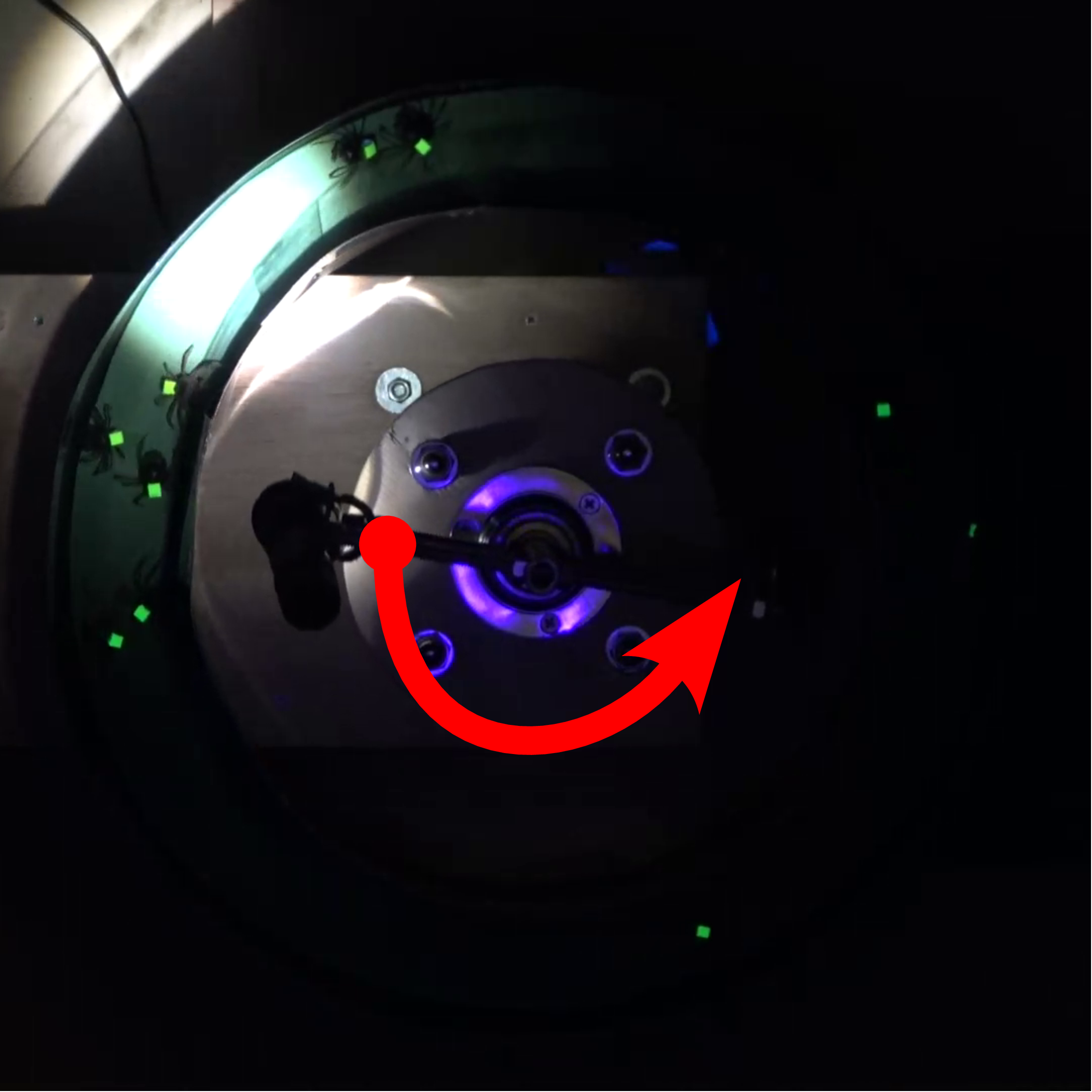}
\caption{Left: The course along which crabs move in the baseline condition. The course is fully lighted and crabs are able to move in each direction. Right: The course is set in darkness and a rotating light is used to induce crabs into moving in the same direction. Ultraviolet light and special markers are used to identify crabs also in darkness.} 
\label{fig:crab_exp}
\end{center}
\end{figure}

Specifically, we want to study whether it is possible to improve the degree of self-organization in swarms of crabs using light (soldier crabs are known to be sensitive to light to some extent). A course similar to the one of the previous experiment is used, but, being 2-3 times in width compared to the size of crabs, it allows motion over multiple lanes in different directions (see Fig.~\ref{fig:crab_exp}). Experiments were performed varying the size of the swarm and were repeated several times. In addition, a simulation model was validated and calibrated to allow obtaining a larger number of data and investigate some aspects not accessible through experimental investigation.

\begin{figure}
\begin{center}
\includegraphics[height=4.0cm]{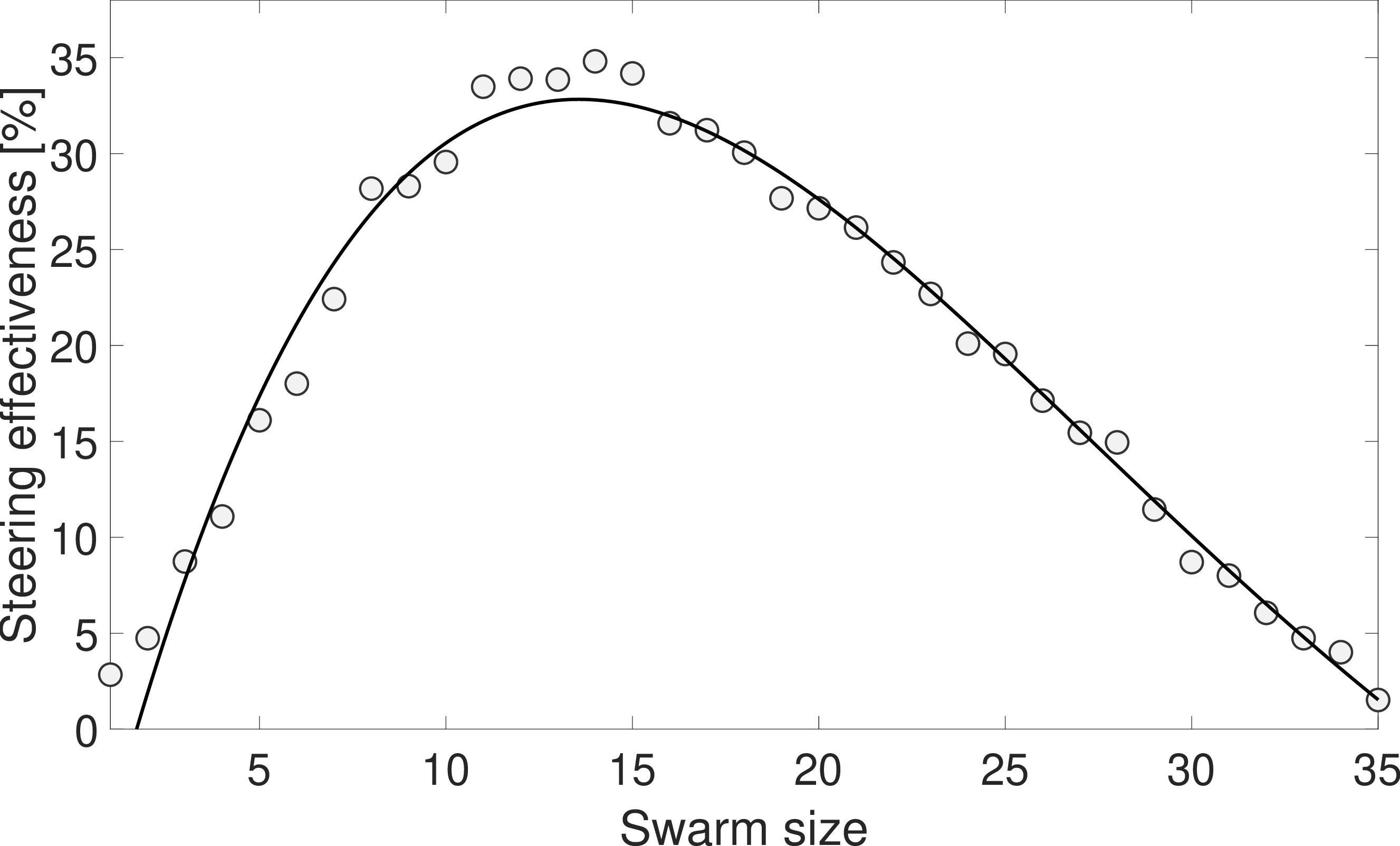}
\caption{Results relative to the capability to improve self-organization of the swarm using the rotating light. Steering effectiveness is obtained by comparing self-organization with the motion observed in the baseline condition. Experimental results are used to validate a simulation model which was used to produce the results presented here (see \cite{Feliciani2022} for details).} 
\label{fig:crab_results}
\end{center}
\end{figure}

Results are presented in Fig.~\ref{fig:crab_results}. The ``steering effectiveness'' measures the extent by which light is capable to improve the self-organization of crabs compared to the baseline without any sort of steering. As it can be seen, effectiveness is low for low densities since individuals are reactive to light only to a limited extent. However, when the swarm gets larger, interactions among the animals contribute to an increase in steering efficiency (the simple combination of multiple interactions is not linear). But the increase in effectiveness is constrained by density, making interactions stronger but also motion more difficult. For this reason, the steering mechanism becomes ineffective at high densities, when crabs are already able to get self-organized without the help of external stimuli.

\section{Conclusion}
In this work, we considered different studies which share similar geometries and all aim at inducing a collective behavior using several control mechanisms. Although some conclusions are specific to each study, there are also common learnings which are shared between them. For instance, density has been commonly found as an important variable determining the effectiveness of steering strategies. This may not seem surprising considering that density is one of the most important properties of crowds (and swarms), but it should remind that it is even more important when control strategies are being developed since some solutions may work well under low densities and others in packed conditions. Compliance was also a common element. It is therefore necessary to consider that crowds may never be fully compliant and be prepared to control them under different degrees of compliance. Finally, we also learned that there is not necessarily a connection between an optimal solution from an engineering perspective and what is considered optimal by the people composing the crowd. At the present stage, automatic control systems designed to work on crowds are still in the development phase and their scope is mostly to deliver guidance in case of emergencies where comfort is only marginal. However, considering that crowds are \textit{always} composed of a physical and a psychological entity, it is also important to start considering the psychological dimension from the early stage of development to ensure the creation of systems well accepted by the public and the main user: the crowd made of people.

\bibliography{ifac2023_feliciani}

\end{document}